# Dimensional reduction, quantum Hall effect and layer parity in graphite films

Jun Yin[1,2], Sergey Slizovskiy[1,2], Yang Cao[1,3,4], Sheng Hu[1,3], Yaping Yang[1,2], Inna Lobanova[5], Benjamin Piot[5], Seok-Kyun Son[1,2], Servet Ozdemir[1,2], Takashi Taniguchi[6], Kenji Watanabe[6], Kostya S. Novoselov[1,2], Francisco Guinea[7,2], A. K. Geim[1,2], Vladimir Fal'ko[1,2], Artem Mishchenko[1,2]

[1]*National Graphene Institute, University of Manchester, Manchester M13 9PL, UK.*
[2]*School of Physics and Astronomy, University of Manchester, Manchester M13 9PL, UK.*
[3]*College of Chemistry and Chemical Engineering, Collaborative Innovation Center of Chemistry for Energy Materials, Xiamen University, Xiamen, 361005, China.*
[4]*State Key Laboratory of Physical Chemistry of Solid Surface, Xiamen University, Xiamen, 361005, China.*
[5]*Laboratoire National des Champs Magnétiques Intenses, LNCMI-CNRS-UGA-UPS-INSA-EMFL, 25 avenue des Martyrs, 38042 Grenoble, France.*
[6]*National Institute for Materials Science, 1-1 Namiki, Tsukuba, 305-0044, Japan.*
[7]*IMDEA Nanoscience, Faraday 9, Cantoblanco, E-28049 Madrid, Spain.*

**The quantum Hall effect (QHE) originates from discrete Landau levels forming in a two-dimensional (2D) electron system in a magnetic field[1]. In three dimensions (3D), the QHE is forbidden because the third dimension spreads Landau levels into multiple overlapping bands, destroying the quantisation. Here we report the QHE in graphite crystals that are up to hundreds of atomic layers thick – thickness at which graphite was believed to behave as a 3D bulk semimetal[2]. We attribute the observation to a dimensional reduction of electron dynamics in high magnetic fields, such that the electron spectrum remains continuous only in the direction of the magnetic field, and only the last two quasi-one-dimensional (1D) Landau bands cross the Fermi level[3,4]. In sufficiently thin graphite films, the formation of standing waves breaks these 1D bands into a discrete spectrum, giving rise to a multitude of quantum Hall plateaux. Despite a large number of layers, we observe a profound difference between films with even and odd numbers of graphene layers. For odd numbers, the absence of inversion symmetry causes valley polarisation of the standing-wave states within 1D Landau bands. This reduces QHE gaps, as compared to films of similar thicknesses but with even layer numbers because the latter retain the inversion symmetry characteristic of bilayer graphene[5,6]. High-quality graphite films present a novel QHE system with a parity-controlled valley polarisation and intricate interplay between orbital, spin and valley states, and clear signatures of electron-electron interactions including the fractional QHE below 0.5 K.**

The Lorentz force imposed by a magnetic field, $B$, changes the straight ballistic motion of electrons into spiral trajectories aligned along $B$ (Fig. 1a). Such spiral motion gives rise to Landau bands that are characterised by plane waves propagating along the $B$ direction but quantised in the directions perpendicular to the magnetic field[7]. The increase in $B$ changes the distance between Landau bands, and the resulting crossings of the Fermi level, $E_F$, with the Landau band edges lead to quantum oscillations. When only the lowest few Landau bands cross $E_F$ (ultra-quantum regime), the magnetic field effectively makes the electron motion one-dimensional (with conductivity allowed only in the direction parallel to $B$). In normal metals such as, e.g., copper, this dimensional reduction would require $B > 10,000$ T. In semimetals with their small Fermi surfaces (e.g., graphite[2,4,8]) the dimensional reduction and the ultra-quantum regime (UQR) can be reached in moderate $B$. Indeed, magnetotransport measurements in bulk crystals[4,9-11] and films of graphite[12,13] revealed quantum oscillations reaching the lowest Landau bands. However, no dissipationless QHE transport could be observed as expected for the 3D systems.

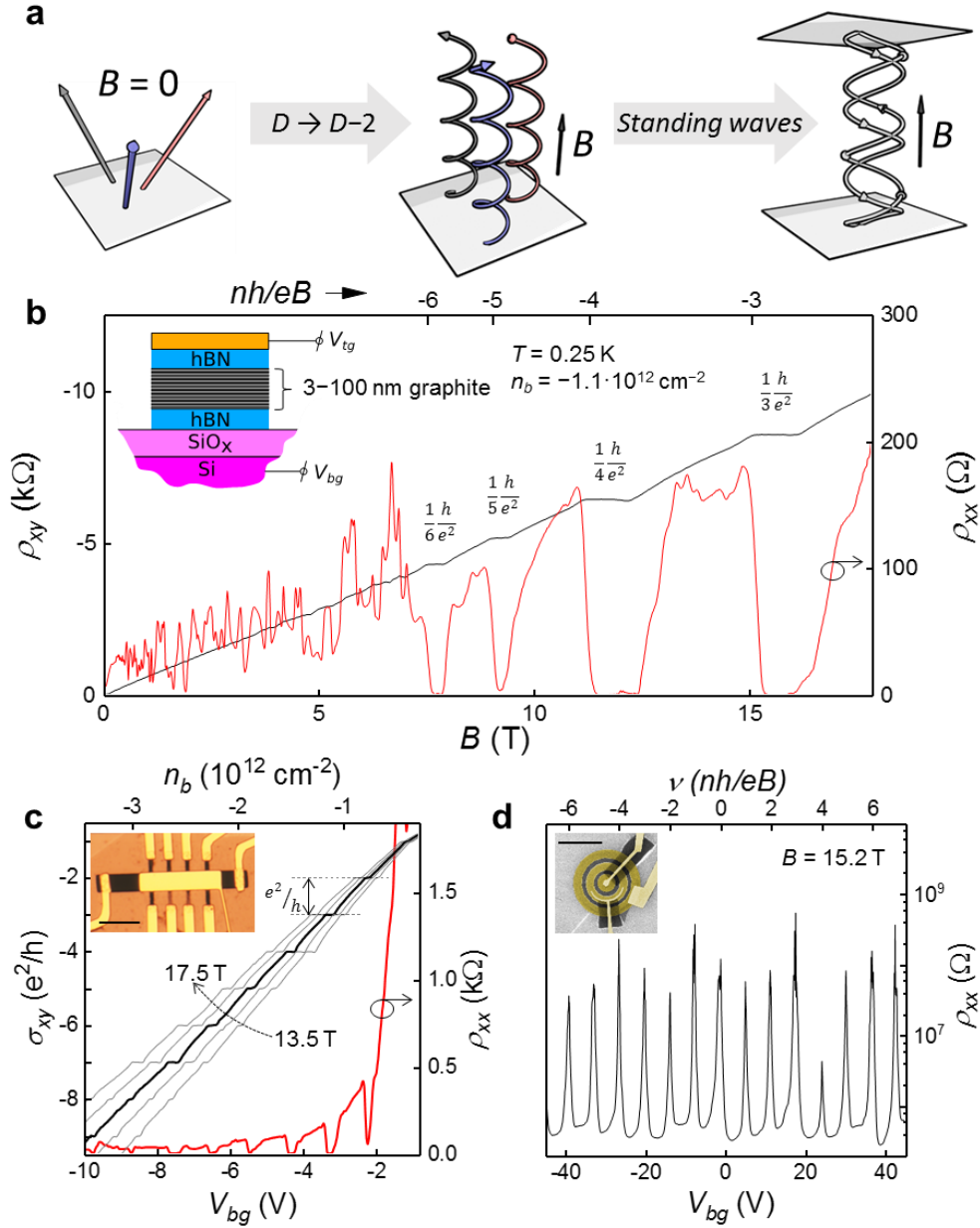

**Fig. 1: Quantum Hall effect in 3D graphite. a,** Ballistic electron trajectories in the absence of magnetic field (left cartoon) undergo dimensional reduction into cyclotron spirals in high fields (middle). In thin crystals, the spiral trajectories can form standing waves (right). **b,** Transversal $\rho_{xy}$ (black curve) and longitudinal $\rho_{xx}$ (red) resistivity as a function of $B$, measured at 0.25 K in a 6 nm thick graphite device. The density $n_b = -1.1\cdot10^{12}$ cm$^{-2}$ is induced by applying a back gate voltage and the negative sign corresponds to holes. The inset shows a schematic of our hBN/graphite/hBN heterostructures. **c,** Hall conductivity $\sigma_{xy}$ (black curve) and $\rho_{xx}$ (red) as a function of back gate voltage $V_{bg}$ at $B = 15.5$ T. Grey curves: measurements at $B$ from 13.5 to 17.5 T; $T = 0.25$ K; $V_{tg} = 0$ V; same device as in **b**. Inset: Optical micrograph of the used Hall bar device (scale bar, 10 μm). **d,** $\rho_{xx}$ in the Corbino geometry; $T$ = 0.25 K; $L$ = 10 nm. $\rho_{xx}$ diverges at the integer filling factors $\nu$ (top x-axis). Inset: False-colour scanning electron micrograph of the Corbino device. Scale bar, 5 μm.

The UQR in graphite can be reached in $B > 7$ T (Ref. 4) where all electrons near the Fermi level belong to the two half-filled 1D bands emerging from the lowest (*0* and *1*) Landau levels of bilayer graphene that is the building block of Bernal-stacked graphite[5]. At the opposite edges of the 3D Brillouin zone (lines +KH and −KH, designated below as valleys) these *0* and *1* states reside on the alternating atomic planes of carbon atoms, and in the bulk graphite crystal the inversion symmetry of its bilayer unit cell makes these spectra valley-degenerate. The 1D character of Landau bands in graphite in the UQR suggests the formation of charge density wave states[14] at $B > 25$ T due to their half-filling and of an insulating phase[15] at even higher $B > 70$ T, where the splitting between the *0* and *1* Landau bands exceeds their bandwidth. In our case of thin graphite crystals, these quasi-1D bands split into a set of standing waves separated by $\Delta k_z = \pi/L$ where $L$ is the crystal thickness (Fig. 1a). The standing waves lead to discretisation of the electronic spectrum and, therefore, can in principle allow the QHE that is otherwise forbidden in 3D. Here we show that, in the UQR, high-quality graphite films of a submicron thickness indeed exhibit the fully developed QHE with quantised Hall and zero longitudinal resistances. Moreover, the standing waves retain the valley-layer correspondence of the *0* and *1* bulk Landau bands so that crystals with an even number of layers $N$ retain the valley degeneracy in the standing-wave spectrum whereas the degeneracy is lifted for odd $N$ because of the lack of inversion symmetry[6]. In a way, the described films exhibit the electronic properties of both 3D and 2D conductors and, hence, can be referred to as a 2.5D system.

To prepare our devices, we cleaved natural graphite crystals along their basal planes using micromechanical exfoliation technique[16]. Then we encapsulated cleaved crystals in hexagonal boron nitride (hBN), expanding the technology of van der Waals heterostructures onto 3D systems[17]. Due to a self-cleansing mechanism, this approach results in quality atomically-flat interfaces with little electron scattering at the surfaces[18,19]. We focus on crystals with thicknesses above 10 graphene layers ($L > 3.5$ nm), which are commonly considered as 3D graphite electronically[20]. For the upper limit of $L$, we chose thicknesses of about 100 nm because an average distance between stacking faults in natural graphite is known to be 100–200 nm[21,22] and those defects efficiently decouple electronic states in different parts of graphene stacks[23]. All our devices showed a metallic temperature dependence of resistivity (see, e.g., Supplementary Fig. 1). Above 30 K the mobility of charge carriers followed a $1/T$ dependence, in agreement with the earlier studies[24,25], and saturated below 10 K to $\sim 4 \cdot 10^5$ cm$^2$/V·s.

As we turned the magnetic field on, our devices exhibited strong positive magnetoresistance such that the longitudinal resistivity $\rho_{xx}$ increased by two orders of magnitude with increasing $B$ from 0 to 0.4 T (Supplementary Fig. 1), which is typical for compensated semimetals[26]. Already at fairly low $B$ of < 0.1 T, we observed clear Shubnikov-de Haas oscillations that developed into the QHE above ~7 T (Fig. 1b). Indeed, one can see that around $B = nh/ev$, the Hall resistivity $\rho_{xy}$ plateaus out at unit fractions of $h/e^2$, which is accompanied by vanishing $\rho_{xx}$. Here, $n$ is the electrostatically controlled carrier density, $h$ the Planck constant, $e$ the elementary charge, and $v$ is the Landau filling factor that relates $n$ to the flux quantum, $h/e$. The QHE is also clearly seen if we fixed $B$ and changed $n$ (Fig. 1c). Note that current densities as low as 50 nA/μm were sufficient to suppress the QHE (Supplementary Fig. 2). Vanishing $\rho_{xx}$ (and $\sigma_{xx}$) signifies dissipationless transport along the edges of the otherwise insulating films[1]. We confirmed these insulating states using the edgeless (Corbino) geometry: Fig. 1d shows the insulating behaviour of a 10 nm thick Corbino device where, for integer $v$, $\rho_{xx}$ exceeds 0.1 GΩ.

In graphite, mobile charge carriers induced by gate voltages are mainly localised in the first couple of layers near the surface because of electrostatic screening. Since the *0* and *1* Landau bands cross $E_F$ at almost half-filling (Fig. 2c), their surface states coexist and mix with the bulk states, making the two graphite surfaces interconnected (correlated). However, for sufficiently high doping the situation is expected to change because higher Landau bands can also become occupied, which results in charge accumulation near the surfaces. The resulting surface states decay exponentially into the bulk, and this breaks down correlations between the two surfaces. To disentangle the contributions from the bulk and surface states to the observed QHE we carried out transport measurements in a double-gated Corbino device (Fig. 2) and electronic compressibility measurements in a double-gated capacitor device (Supplementary Fig. 3). With the two gates, we can induce charge carriers at the top ($n_t$) and bottom ($n_b$) surfaces independently. Longitudinal conductivity $\sigma_{xx}(B)$ as a function of the electrostatically-induced density $n_{tot} = n_t + n_b$ is shown as a colour map in Fig. 2a where metallic states (bright colours) fan out from the neutrality point, $n_{tot}$ = 0. The majority of the observed features can be attributed to the discussed 2.5D QHE states as in high fields they are linear in $B$, and their slopes $B/n_{tot}$ follow integer filling factors. However, there is a notable change at ~7 T, below which the QHE is obscured by a complicated criss-crossing behaviour (the existence of the critical field manifests itself even more clearly in thicker devices, see Supplementary Fig. 4). In addition, we also observe features that are not linear in $B$ and do not follow integer $v$, examples of which are indicated by the dashed curves in Fig. 2a. We attribute those quantised levels to the 2D accumulation layers at the two graphite surfaces (for discussion, see Methods). The difference between the quantised features due to the surface states and those due to the 2.5D QHE is especially obvious from maps $\sigma_{xx}$ ($n_t$, $n_b$) at a fixed $B$ (Fig. 2b; also, Supplementary Fig. 3). The 2.5D QHE states formed by the standing waves in the bulk follow the diagonal lines on the map, that is, both top and bottom gates contribute equally to the filling of the Landau states that penetrate through the whole bulk. On the contrary, the quantised states at the top and bottom surfaces follow horizontal and vertical lines, respectively, which indicates that the states are electronically decoupled and interactions between the two surfaces are screened out[27]. To better understand the origin of the surface states we performed self-consistent tight-binding/Hartree calculations (Methods), and the computed density of states along with carrier density profiles are presented in Supplementary Fig. 3.

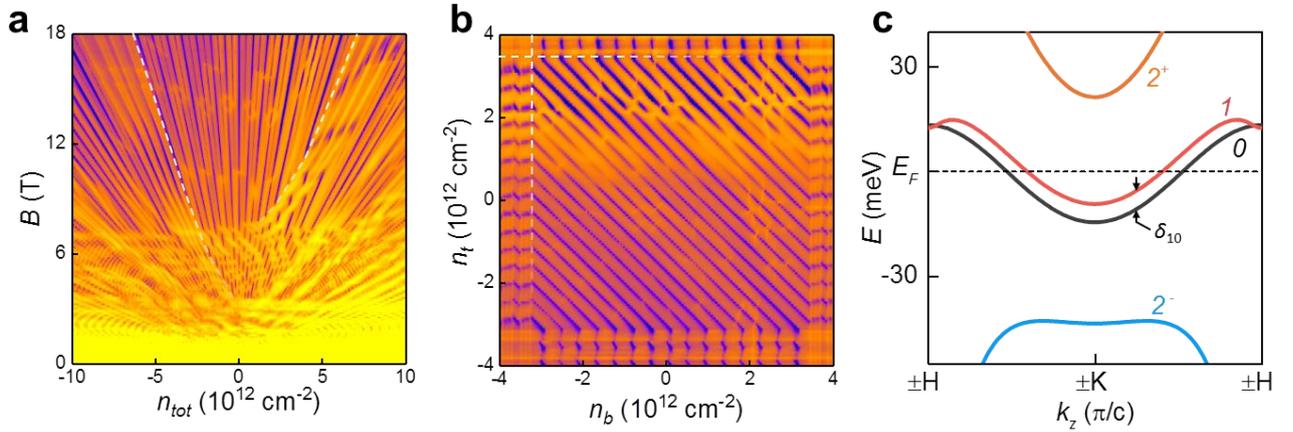

**Fig. 2: Landau levels in thin graphite films. a,** $\sigma_{xx}$ as a function of $B$ and $n_{tot}$ measured at $n_t = n_b$ in a Corbino device with $L$ = 11.6 nm; $T$ = 0.25 K. The white dashed curves are guides to the eye for the features arising from the Landau quantisation of electrostatically induced surface states. **b,** Map $\sigma_{xx}$ ($n_t$, $n_b$) at 18 T. The horizontal and vertical features at the edges of the map (contoured by white lines) are due to the quantised states at the graphite surfaces. The diagonal features are ascribed to the $k_z$-quantised lowest Landau bands that extend through the graphite bulk (2.5D QHE). Logarithmic colour scales for both **a** and **b**; navy-to-yellow is 50 nS to 50 μS. **c,** Dispersion relation for low Landau bands calculated using the tight-binding model at 15 T (Methods). Only the lowest two bands (labelled as *0* and *1*, and split by $\delta_{10}$) cross the Fermi level.

The conductivity maps in Fig. 2 also show that the gaps along the integer filling factors can close and reopen as a function of both $n$ and $B$, which is indicative of level crossings. To explain the observed patterns, we refer to the band structure of graphite in the Slonczewski-Weiss-McClure tight-binding model[2,8,28,29] for Bernal-stacked graphite (see Methods). At zero field, the low energy electronic structure is dominated by two bands in two valleys[2,8,28] that are degenerate for $k_{x,y}$ = ±K and weakly disperse in $k_z$ (+KH and −KH valleys). In quantising fields[4,29], these are reduced to two nearly coinciding Landau bands *0* and *1* (Fig. 2c). They cross the Fermi level close to $k_F$ ≈ π/4$c$ ($c$ = 0.335 nm is the interlayer distance in graphite), exhibit the 1D Fermi velocity $v_F$ ≈ 15,000 m/s, and are split by $\delta_{10}$ ≈ 0.37 meV/T (Methods).

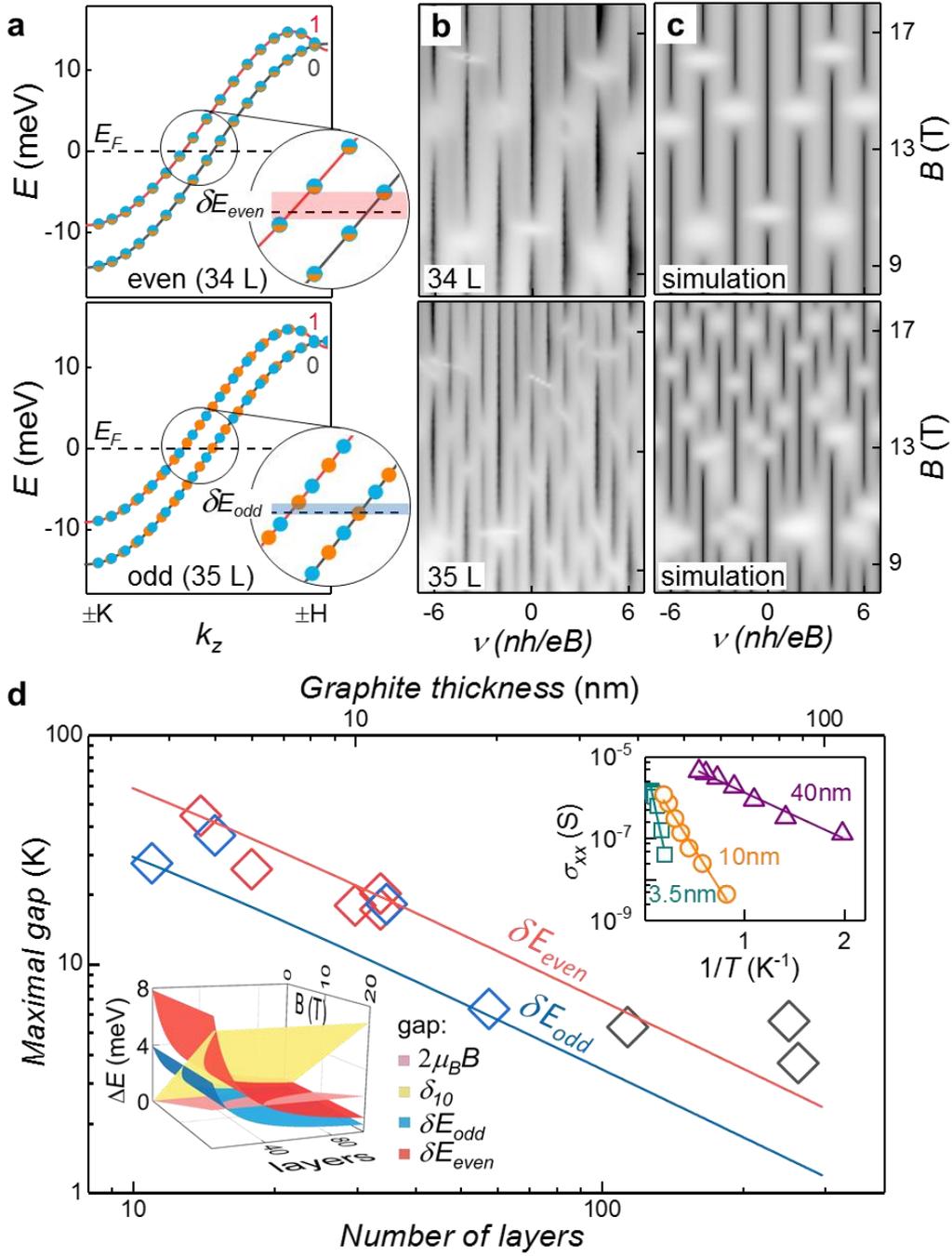

**Fig. 3: Thickness and layer-parity dependence of the 2.5D QHE. a,** $E(k_z)$ for the two lowest Landau bands in graphite (grey and red curves). Blue and orange dots are the standing wave states caused by $k_z$-quantisation calculated for +KH and −KH valleys, respectively. The top and bottom panels are for even $N = 34$ and odd $N = 35$, respectively. **b,** $\sigma_{xx}(B, \nu)$ for 34 and 35 layer devices (top and bottom panels, respectively) measured at 1 K. Logarithmic grey scale: black-to-white is 0.1 to 5.6 µS. Note the double degeneracy ($\Delta\nu = 2$) for even-layered graphite and the fully lifted degeneracy ($\Delta\nu = 1$) for the odd-layered one. **c,** Calculated density of states as a function of $B$ and $\nu$ for the graphite device in **b**. Logarithmic scale: black-to-white is 0.01 to 1.5 eV$^{-1}$nm$^{-2}$ and 0.01 to 1 eV$^{-1}$nm$^{-2}$ for the top and bottom panels, respectively. **d,** Energy gaps for the 2.5D QHE as a function of thickness. Shown are the maximal gaps found away from level crossings. Red symbols – graphite devices with even $N$, blue symbols – odd $N$, black – unconfirmed parity. The red and blue lines are the gaps estimated from the free-particle model. Top inset: examples of Arrhenius plots for three devices. Bottom inset: relative strengths of $\delta E_{even/odd}$, $\delta_{10}$ and $2\mu_B B$ (Zeeman splitting) gaps as a function of $N$ and $B$.

In graphite crystals with a small *L*, electrons in the two Landau bands *0* and *1* form standing waves, which for different valleys reside on graphene layers of the different parity: the states in +KH valley live on the even layers, and the states in −KH valley on the odd ones (see Methods). For even-*N* crystals, there is the same number of even and odd layers to support states in both +KH and −KH valleys, leading to the valley-degenerate ladder of states that are spaced in $k_z$ by the distance, $\frac{\pi}{c(N+2)}$. In odd-*N* crystals, the number of even and odd states differs by one, which results in a $k_z$-shift of $\approx \frac{\pi}{2c(N+1)}$ between the two ladders, around the Fermi level. This shift lifts the valley degeneracy and, therefore, reduces the energy spacing, $\delta E_{odd}$, nearly twice as compared to the even-*N* films (see Fig. 3a). We refer to this difference between the standing wave spectra in graphite with even and odd *N* as the layer-parity effect.

To test for this effect, we measured the 2.5D QHE in several pairs of Corbino devices, in which the thickness differed only by a single graphene layer. We fabricated these pairs from same graphite crystals having monoatomic steps on the surface (Methods). The transport measurements were performed in sufficiently high *B* to induce spin polarisation, that is, $2\mu_B B > k_B T$ where $\mu_B$ is the Bohr magneton and $k_B$ the Boltzmann constant. Fig. 3b shows that for the 34-layer graphite device the degeneracy of the QHE states was 2 whereas all the unit steps in *v* were observed for the device with *N* = 35. This indicates that all the degeneracies were lifted in the latter case, including the valley degeneracy. Data for devices with *N* = 14 and 15 are presented in Supplementary Fig. 5.

The pronounced level crossings seen in the maps of Figs. 2 and 3 can also be related to specifics of the 2.5D QHE in graphite. This behaviour arises due to competition between the Landau band splitting, $\delta_{10}$, the spin splitting, $2\mu_B B$, and the *B*-independent standing waves spacing, $\delta E_{even/odd}$, which all lead to a complex hierarchy of energy gaps and numerous level crossings with changing *B*. The bottom inset in Fig. 3d summarises the expected interplay between the gap sizes for different fields, thicknesses and parities (Methods and Supplementary Fig. 6). Figure 3 also shows the measured maximal QHE gaps, *ΔE*, for each of the studied samples with different *L*. The gaps were measured away from level crossings using the Arrhenius model for the dissipative conductivity in the QHE regime, $\sigma_{xx} \propto e^{-\Delta E/2k_B T}$ (top inset of Fig. 3d). The red and blue curves show the upper bounds for *ΔE* as found in our modelling for even and odd *N*, respectively.

The level crossing behaviour is further detailed in Fig. 4. Figure 4a compares the measured and simulated $\sigma_{xx}$ (*B*, *v*) for the 11.6 nm device in Fig. 2. In both cases there are two periods, *Δv*, at which crossings occur: *Δv* = 4 and 8. For *N* > 20, we estimate (see Methods) that $\delta E_{even/odd} < \delta_{10}$ for the experimental range of *B* and, therefore, the level crossings are dominated by fast changes in the Landau band splitting. Figure 4b highlights the complex hierarchy of gaps for different *B*. For example, at about 13 T, four valley-degenerate levels are coming close to each other, which effectively results in relatively large gaps between crossings and, therefore, in robust quantisation such that steps with *Δv* = 8 persist to $k_B T \approx \delta E_{even} \approx 94\ N^{-1}$ meV (this dependence is also shown by the red curve in Fig. 3d). On the other hand, at ∼ 10 T, valley-degenerate levels cross at some distance from each other, resulting in approximately twice smaller gaps and steps with *Δv* = 4, as observed in both experiment and simulations. Similarly, for films with odd *N*, in which the valley degeneracy is lifted, the upper limit for the 2.5D QHE gaps is set by $\delta E_{odd} \approx 47\ N^{-1}$ meV (blue curve in Fig. 3d). Note that the observed dependences *δE*(*N*) imply that, at mK temperatures, the 2.5D QHE should be observable even in micron-thick graphite, if of course no stacking faults are present which would probably suppress the quantisation.

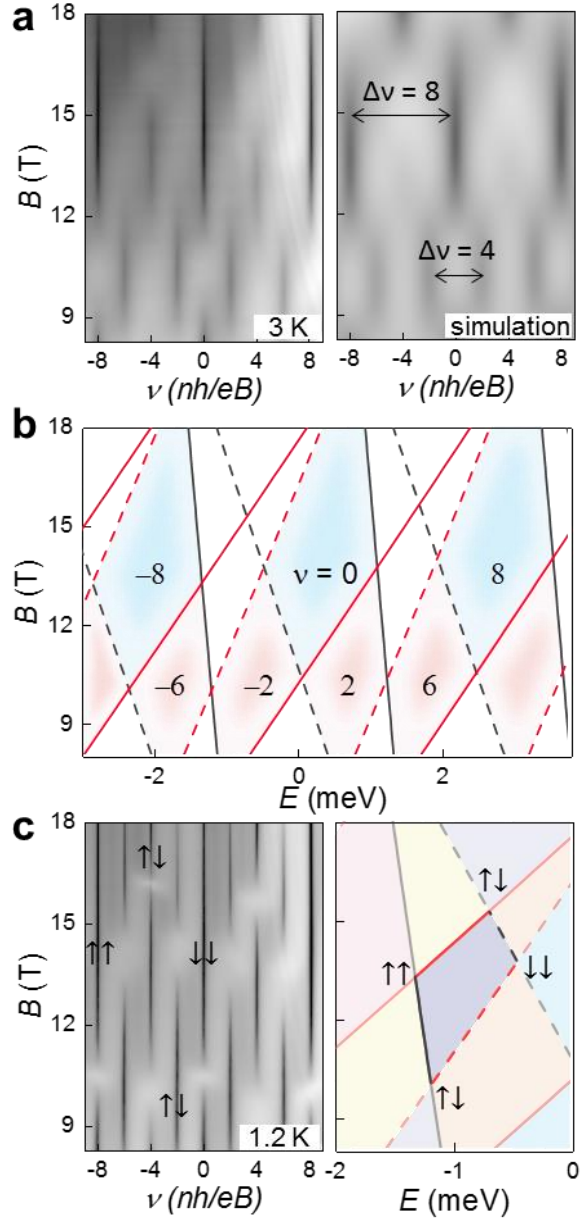

**Fig. 4: Hierarchy of QHE gaps and level crossings in the ultra-quantum regime. a,** Left panel: Map $\sigma_{xx}(B,\nu)$ for the device in Fig. 2; $T = 3$ K. Logarithmic grey scale: black-to-white is 0.6 to 5.6 µS. Right panel: Calculated density of states as a function of $B$ and $\nu$. Colour scale: black-to-white is 0.1 to 2 eV$^{-1}$ nm$^{-2}$. **b,** Calculated $0$ (black) and $1$ (red) Landau bands of 34-layer graphite. Solid (dashed) lines are ↑ (↓) spin states. The Landau level crossings result in QHE-gap closures in **a**. **c,** Left panel: same as in the left panel of **a** but the measurements were done at 1.2 K. Logarithmic scale: black-to-white is 0.1 to 7.4 µS. Right panel: Zoomed-in region of **b**. Gaps closures in the left panel of **c** and the level crossings in the right panel of **c** are labelled by the corresponding spin states ↑ and ↓. The coloured background specifies $\nu = -8$ (pink), $-6$ (yellow), $-4$ (violet), $-2$ (orange), and $\nu = 0$ (blue).

The free-particle (non-interacting) model describes well all the main characteristics of the observed QHE confirming that the phenomenon arises due to the standing waves and is controlled mostly by graphite's thickness and layer parity. However, at our highest $B$ and lowest $T$, the devices also exhibit features indicating that electron-electron interactions clearly contribute to the electronic properties. For example, Fig. 4c and Supplementary Fig. 7a show that crossings between the same-spin states are much broader than those between the opposite-spin states, which suggests some magnetic ordering[30]. Besides, even-layer graphite devices exhibit spontaneous valley polarisation at $T < 0.5$ K (Supplementary Fig. 7b-g). For relatively thin devices ($L$ = 3-4 nm) we also observed the onset of the fractional QHE (Supplementary Fig. 8). It is generally expected that electron-electron interactions can trigger a variety of phase transitions in the form of density waves, excitonic condensation or Wigner crystals[15,31,32], and the high-quality graphite films introduced in this work offer an enticing playground (especially, at mK temperatures) to study such transitions with many competing order parameters.

# Methods
## Device fabrication
Our hBN/graphite/hBN heterostructures were assembled using the dry transfer technology[17,33]. Graphite and hBN flakes were first mechanically exfoliated onto oxidised Si wafers. Then the top hBN and graphite crystals were sequentially picked up by a polymer bilayer of polydimethylsiloxane (PDMS) and polymethylmethacrylate (PMMA). The hBN/graphite stack was released onto the bottom hBN crystal which completed the heterostructure. It is known that accurate crystallographic alignment between graphite and hBN crystal lattices can result in superlattice effects and, therefore, we intentionally misaligned graphite with respect to both hBN crystals by angles larger than 5°. This makes the moiré period small and shifts the band structure reconstruction to energies inaccessible in transport measurements. Raman spectroscopy was also employed to confirm that the chosen graphite crystals had the Bernal (ABA) stacking[34] (Supplementary Fig. 9).

Electrical contacts to graphite were prepared by e-beam lithography, which created openings in a PMMA resist and allowed us to remove the top hBN layer using reactive ion etching. E-beam evaporation was then used to deposit metal contacts Cr/Au (3 nm/80 nm) into the etched areas. Another round of e-beam lithography and e-beam evaporation was implemented to define the top gate, which also served as an etch mask to define the Hall bar geometry. In the case of Corbino devices, crosslinked PMMA made by overexposure of the e-beam resist was used as an insulating bridge between the inner and outer contacts.

To investigate the layer parity effect, pairs of Corbino devices were fabricated on same graphite flakes that were chosen to have isolated monoatomic steps on the surface. The selection of graphite was based on the optical contrast but the step height was subsequently checked by atomic force microscopy. We then determined the layer parity of the graphite at each side of the step based on the degeneracy period of the fan diagram, which, in combination with measured thickness, allowed us to unambiguously identify the number of layers.

In total, we studied approximately twenty devices in both Hall bar and Corbino geometries with thicknesses ranging from 3.5 to 93 nm. In addition, to distinguish between surface and bulk effects, we fabricated several devices on quartz substrates to carry out capacitance measurements. Furthermore, to check whether stacking faults in graphite destroy the observed 2.5D QHE, we prepared special hBN/graphite-graphite/hBN heterostructures in which the graphite film consisted of two crystals that were intentionally slightly misaligned to form an artificial stacking fault. Such devices exhibited no insulating states essential for observation of the QHE (Supplementary Fig. 9a,b). We also note that hBN encapsulation is crucial for the observation of the 2.5D QHE as witnessed in experiments using similar devices but with graphite placed in direct contact with the rough quartz substrate (no bottom hBN). Its absence ruined the electronic quality of the bottom graphite surfaces, which was sufficient to smear the 2.5D QHE gaps completely (Supplementary Fig. 9d,e). On the other hand, for the properly encapsulated graphite the 2.5D QHE was quite robust and observed in all our devices, even for lateral sizes up to tens of micrometres (limited only by the maximum area we could find that was free from contamination bubbles) (Supplementary Fig. 9f).

## Transport and capacitance measurements
The longitudinal and Hall voltages were recorded using lock-in amplifier SR830 and applying ac currents typically of about 10 nA. Such unusually small currents were found necessary to avoid the suppression of the observed 2.5D QHE (Supplementary Fig. 2). For measurements on the Corbino devices, a small bias (typically 0.1 mV) was applied between the inner and outer contacts, and the current was recorded using again an SR830 amplifier. The capacitance was measured using an on-chip cold bridge based on a high-electron mobility transistor[35]. The typical excitation voltages were in the range from 1 to 10 mV, depending on the thickness of the hBN dielectric layer.

Away from the QHE regime (high temperatures and low magnetic fields), $\sigma_{xx}$ and $\sigma_{xy}$ in semimetallic graphite can be described by the expressions $\sigma_{xx}(n_i, \mu_i, B) = \sum_i \frac{n_i e \mu_i}{1+\mu_i^2 B^2}$ and $\sigma_{xy}(n_i, \mu_i, B) = \sum_i \frac{n_i e \mu_i^2 B}{1+\mu_i^2 B^2}$ where $n_i$ and $\mu_i$ are the carrier densities and mobilities, respectively, and index $i$ runs through different types of charge carriers. Using the above expressions, we fitted the measured $B$ and $T$ dependences for $\sigma_{xx}$ and $\sigma_{xy}$ and found that the model with just two types of carriers provides good agreement with the experiment. This also allowed us to estimate the charge densities and mobilities in the absence of electrostatic doping (Supplementary Fig. 1).

*Tight-binding description of Landau bands in the ultra-quantum regime*

We employed the Slonczewski-Weiss-McClure (SWMC) model[2,8,29] with hopping amplitudes, $\gamma_0$ = 3.16 eV, $\gamma_1$ = 0.39 eV, $\gamma_2$ = −17 meV, $\gamma_3$ = 0.315 eV, $\gamma_4$ = 70 meV, $\gamma_5$ = 38 meV, and the dimer site energy shift, $\Delta$ = −5 meV. While most of these values have been well established in the earlier studies, there is no clear consensus about the values of $\gamma_2$, $\gamma_5$ and $\Delta$, hence, we adjusted the latter to get the best fitting to the experimental results obtained in the present work. Namely, we chose $\gamma_2$, $\gamma_5$ and $\Delta$ to match (a) the slope of *0, 1* Landau bands dispersion near the Fermi level which is given by $\gamma_2$ with an extra correction coming from $\gamma_3$ and (b) the distance $\delta_{10}$ between *0* and *1* Landau bands.

The zero field spectrum of graphite, determined by the SWMC model, contains 4 bands. Two high-energy bands, $\epsilon_\pm^{\text{split}}(k_z, p)$,

$$\epsilon_\pm^{\text{split}}(k_z, p) = \frac{1}{2}(E_\pm + E_3) \pm \sqrt{\frac{1}{4}(E_\pm - E_3)^2 + \left(1 \mp \frac{\gamma_4 \Gamma}{\gamma_0}\right)^2 (vp)^2},$$

form a minority hole Fermi surface at the H point ($k_z = \pi/2c$) whereas the other two degenerate low-energy bands $\epsilon_\pm^{\text{low}}(k_z, p)$, reminiscent of the bilayer graphene low-energy bands[5],

$$\epsilon_\pm^{\text{low}}(k_z, p) = \frac{1}{2}(E_\mp + E_3) \pm \sqrt{\frac{1}{4}(E_\mp - E_3)^2 + \left(1 \pm \frac{\gamma_4 \Gamma}{\gamma_0}\right)^2 (vp)^2},$$

form majority hole and electron Fermi surfaces. Here, $v = \frac{\sqrt{3}\gamma_0 a}{2\hbar}$, $p = \sqrt{p_x^2 + p_y^2}$, $E_3 = \frac{1}{2}\gamma_2 \Gamma^2$, $E_\pm = \Delta \pm \gamma_1 \Gamma + \frac{1}{2}\gamma_5 \Gamma^2$, $\Gamma = 2 \cos ck_z$, $a$ = 0.246 nm, $c$ = 0.335 nm, and $\hbar = h/2\pi$.

In a magnetic field oriented along the $z$-axis, the in-plane momentum ***p*** has to be extended using Kohn-Luttinger substitution, $\boldsymbol{p} \to -i\hbar\nabla + e\boldsymbol{A}$, and this leads to the Landau quantization of the lateral motion of electrons and formation of Landau bands[26,34,35]. For $\gamma_3 \to 0$ there is an analytic expression[2,8,29,36,37] for valley-degenerate (the symbol ± below is not affected by the choice of the valley) both high-energy ('split') bands,

$$\epsilon_\pm^{\text{split}}(k_z, m) \approx E_\pm(k_z) + eB\hbar v^2 (2m-1)\omega_\pm, \text{ (}m\text{ = 2, 3, ...), } \omega_\pm = \frac{\left(1 \mp \frac{\gamma_4 \Gamma}{\gamma_0}\right)^2}{E_\pm(k_z) - E_3(k_z)},$$

and the low-energy bands ($m$ = 2, 3, ...),

$$\epsilon_\pm^{\text{low}}(k_z, m) \approx E_3(k_z) - eB\hbar v^2 \left[\left(m - \frac{1}{2}\right)(\omega_+ + \omega_-) \mp \sqrt{\left(m - \frac{1}{2}\right)^2 (\omega_+ - \omega_-)^2 + \omega_+ \omega_-}\right].$$

Additionally, the two lowest energy bands ($m$ = 0, 1) are described by

$$\epsilon_0(k_z) = 2\gamma_2 \cos^2 ck_z \quad \text{and} \quad \epsilon_1(k_z) = 2\gamma_2 \cos^2 ck_z - eB\hbar v^2(\omega_+ + \omega_-)$$

for which the wavefunctions are localised at even (odd) layers in +KH (−KH) valleys. In magnetic fields $B$ > 7 T, the large distance to the higher Landau bands drives graphite into the UQR, leaving only $\epsilon_{0,1}(k_z)$ bands to cross the Fermi level at $k_F \approx \pi/4c$. This yields the 1D Fermi velocity along the $z$-axis, $\hbar v_{Fz} \approx -2\gamma_2 c$ for the electron motion across the film, and the splitting energy between *0* and *1* Landau bands

$$\delta_{10} = \varepsilon_1 - \varepsilon_0 \approx eB\hbar v^2 \left[\frac{\gamma_5 + \Delta}{\gamma_1^2} + \frac{4\gamma_4}{\gamma_0 \gamma_1}\right] \approx 6.15 \mu_B B,$$

where $\mu_B = \frac{e\hbar}{2m_e}$ is the Bohr magneton. For the parameters listed above, we estimate $\delta_{10}$ ≈ 0.37 meV/T, which is about 3 times larger than the single-particle Zeeman energy for electrons in graphite. The Fermi velocity $v_{Fz}$, and the splitting $\delta_{10}$ are the parameters that we have matched to the measured gaps and level-crossings to fix the above-quoted values of tight-binding model parameters $\gamma_2$, $\gamma_5$, and $\Delta$. In Supplementary Fig. 6, we compare the values of $\delta_{10}$ to the Zeeman splitting and the standing-wave level spacing in graphite of different thickness and layer parity.

Note that graphite spectrum is not particle-hole symmetric because of the non-zero value of hopping parameter $\gamma_2$. This is seen in Fig. 2a and Supplementary Fig. 3a, where the experimental maps do not exhibit electron-hole symmetry. For instance, the localised surface states of electron-hole asymmetric Landau bands $2^+$ and $2^-$ (see Fig. 2c), which are responsible for the surface accumulation layers and screening of the gate potential, are reached by the Fermi level and get filled in different $B$ for the same level of n- and p-doping.

*Standing waves in the ultra-quantum regime and the layer parity effect*

For our thin graphite devices, it has been found that measurable QHE gaps start to appear in $B$ as low as ~0.5 T (Supplementary Fig. 4). However, for crystals with hundreds of atomic layers, the gaps become sufficiently large, clearly visible and regularly structured only for $B > 7$ T (in the UQR), where only the 0 and 1 Landau bands, $\epsilon_{0,1}$, are left to cross the Fermi level. The spectrum of such films can be described using $k_z$ standing waves within the quasi-1D bands $\epsilon_{0,1}$ and the allowed energies are determined by size quantisation of $k_z$. Electrons in the two Landau bands *0* and *1* form standing waves with the maxima on even graphene layers for valley +KH and on odd layers for valley –KH (swapping valleys for $B \rightarrow -B$). For a crystal with an even number of layers, $N = 2\mathcal{N}$, there is the same number of even and odd layers to support states in +KH and –KH valleys, which determines the valley degenerate standing waves with the amplitude $\sin(\frac{\pi n j}{\mathcal{N}+1})$ on the *j*-th graphene layer among even or odd sequences. This corresponds to wave numbers $k_n = \frac{\pi n}{2c(\mathcal{N}+1)}$ where *n* = 1…$\mathcal{N}$, and Landau band energies $\varepsilon_0(k_n)$ and $\varepsilon_1(k_n)$. In graphite with an odd number of layers, $N = 2\mathcal{N}-1$, there are $\mathcal{N}$ odd and $\mathcal{N}-1$ even layers and the standing waves are such that for the valley –KH (on odd layers) $k_n = \frac{\pi n}{2c(\mathcal{N}+1)}$ whereas for valley +KH (on even layers) $k'_n = \frac{\pi n}{2c\mathcal{N}}$. The difference between standing-wave spectra with even and odd *N* is illustrated in Fig. 3a and referred to as the layer parity effect.

Near the Fermi-level ($n_F \approx \mathcal{N}/2$) quantised states in each valley form a staircase of levels separated by $\frac{\pi v_F}{2\mathcal{N}c} \approx \frac{\pi|\gamma_2|}{\mathcal{N}}$. For even *N*, these spectra are valley-degenerate and their spacing is given by $\delta E_{even} \approx \frac{\pi|\gamma_2|}{\mathcal{N}} \approx 94N^{-1}$ meV whereas for odd *N*, the energy states in valley +KH are shifted by $\frac{\pi v_F}{4\mathcal{N}c}$ with respect to those in valley –KH. This lifts the valley degeneracy and, also, reduces the level spacing to $\delta E_{odd} \approx \frac{\pi|\gamma_2|}{2\mathcal{N}} \approx 47N^{-1}$ meV. These energy spacings should be compared to the Landau band splitting given by $\delta_{10} \approx 0.37$ meV/T, and the Zeeman splitting, $2\mu_B B$ because altogether they determine the gaps in the quantised spectrum and the hierarchy of 2.5D QHE plateaux.

In the case of even *N*, valley degeneracy may be broken by an electric field applied in the *z*-direction (Supplementary Fig. 7). However, this effect should disappear in symmetrically gated devices ($n_t = n_b$) due to the recovery of inversion symmetry. For odd-*N* devices, the symmetric gating would still lead to corrections in the valley splitting, because states in Landau bands *0* and *1* are located on the first and last layer in one valley (and, therefore, are stronger influenced by gating) whereas states in the other valley are located on the second and penultimate layers and, therefore, are weaker affected by the same gating. Such corrections can explain a noticeable difference between the measured and simulated fan diagrams for the 35-layer graphite device, Fig. 3b,c, as well as the deviations of the measured gap energies for odd-layer graphite in Fig. 3d from the theoretical curve. In the latter case, for the two blue points that fall onto the red line, the maximal gaps were found at large filling factors. Because for odd *N* gate-induced energy shifts of nearby states in different valleys are different the simple free-particle model is expected to work well only for the first few filling factors where the gate-induced electric field is small. Therefore, notable deviations from the model predictions for these two devices are not surprising.

When we discuss the 2.5D QHE we use the analogy between the Landau level states in bilayer graphene[5] and the states in the lowest two Landau bands in the UQR graphite. That is, we attribute an $e^2/h$ contribution towards $\sigma_{xy}$ from each filled standing wave (each quantised $k_z$) of *0* and *1* Landau bands in a given spin and valley state, counting n-type and p-type doped filling from the neutrality condition, $n_{tot} = 0$. As a result, $\sigma_{xy}$ at integer filling factors is exactly proportional to the charge density induced electrostatically in graphite: $\sigma_{xy} = \nu e^2/h = n_{tot} e/B$, in agreement with the case of the conventional QHE.

*Surface states*

To understand the additional features in the observed Landau fan diagrams (Fig. 2a and Supplementary Fig. 3a), related (as indicated by Fig. 2b and Supplementary Fig. 3b) to the formation of surface states, we appended the SWMC model by the self-consistent analysis of gate-voltage-induced surface charge accumulation, implemented in the Hartree approximation (see Ref. 38). In Supplementary Fig. 3c and 3d we show examples of numerically stable self-consistent solutions for potential and charge distributions on few surface layers for several values of gate-induced charge carrier densities $n_t$. For large $n_t$, one can see in Supplementary Fig. 3c the regions of enhanced compressibility (higher density of states) which correspond to filling localised surface states that split-off from the higher ($m = 2^{\pm}$ and $3^{\pm}$) Landau bands of bulk graphite with electrons trapped in a surface accumulation layer by the self-consistent potential $U(z)$, Supplementary Fig. 3d. In particular, the $2^+$ Landau band forms a surface state for electrons that spawns from the neutrality point at $B \approx 7.5$ T in experimental data (Supplementary Fig. 3a) and at $B \approx 6$ T in theoretical modelling (Supplementary Fig. 3c). Similarly, the localised states chipping off the higher-energy Landau band $3^+$ on the n-doping side, and also the Landau band $2^-$ on the p-doping side can be identified at smaller $B$ in Supplementary Fig. 3.

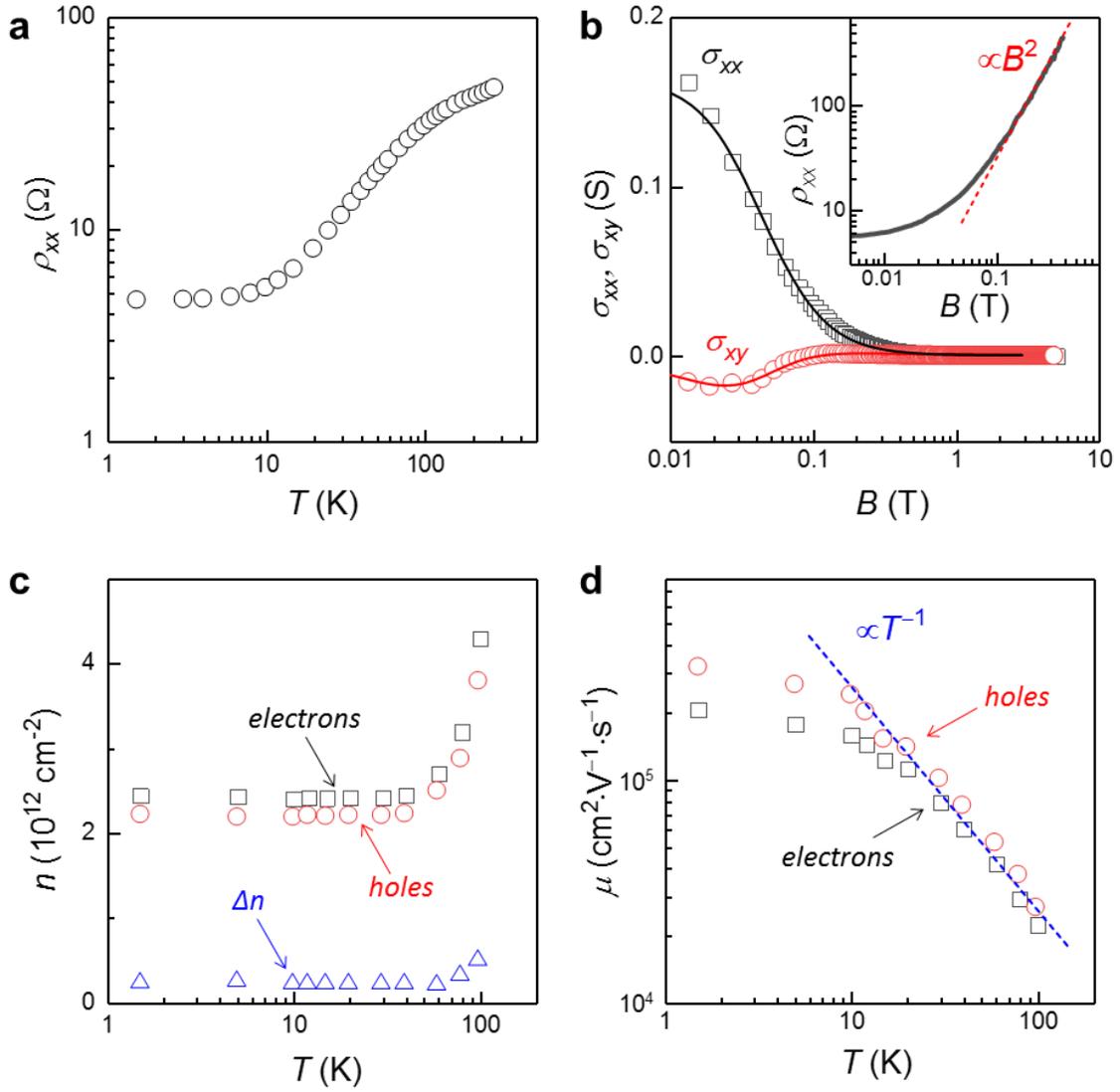

**Supplementary Fig. 1: Transport properties of a 6 nm thick graphite device. a,** Temperature dependence of $\rho_{xx}$ at zero gate doping at $B = 0$ T. $\rho_{xx}$ decreases 10 times (from about 46 to 4.6 Ω) from room to helium temperature. **b**, Measured $\sigma_{xy}$ and $\sigma_{xx}$ as a function of $B$ at 5 K (symbols). The solid curves are best fits using the two-carrier model (see section Transport and capacitance measurements in Methods). The inset illustrates the strong positive magnetoresistance measured at 0.25 K, which above 0.1 T follows the quadratic dependence, in agreement with the fact that graphite is a nearly-compensated semimetal. **c** and **d,** Charge carrier densities and mobilities, respectively, at different $T$. The values were extracted using the two-carrier model as in **b**. At high temperatures, the mobilities follow the $1/T$ dependence.

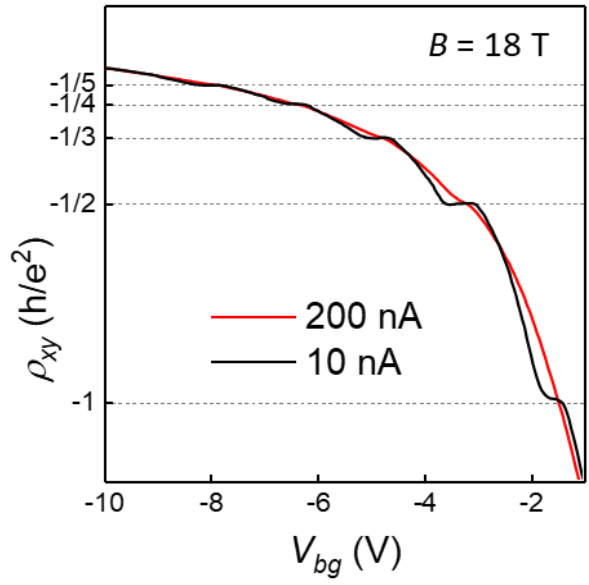

**Supplementary Fig. 2: Effect of high excitation currents on the observed 2.5D QHE.** $\rho_{xy}$ at 0.25 K in a graphite Hall bar device with $L$ = 8 nm, which has the width of 4 µm.

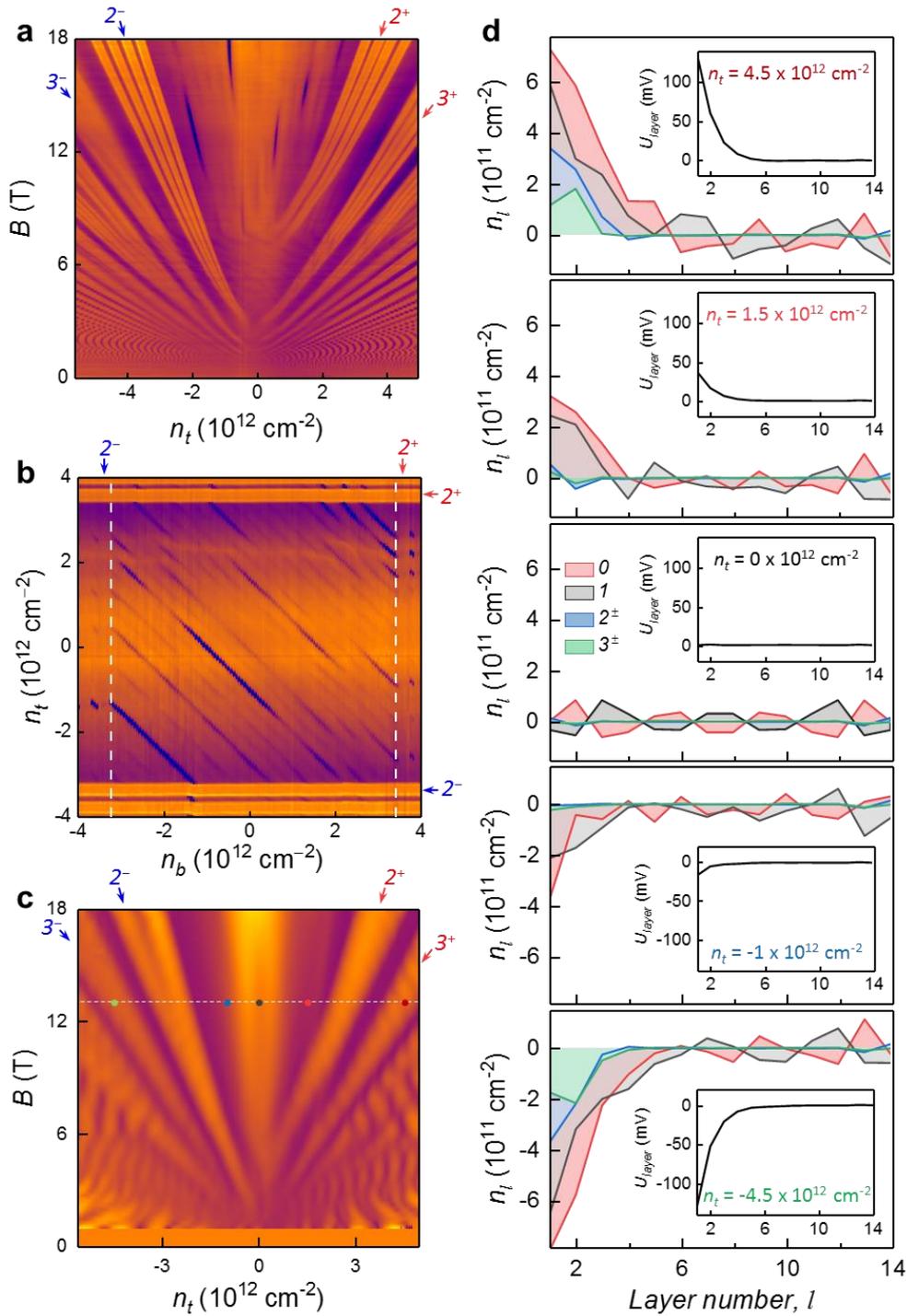

**Supplementary Fig. 3: Surface states and charge carrier distribution in the UQR. a,** Capacitance $C$ as a function of $n_t$ and $B$, measured for a 20 nm thick graphite device at 0.25 K. **b,** Map $C(n_b, n_t)$ for the same device at $B = 18$ T. The gaps (dark colours) slope at 45°, indicating that they depend on the total doping $n_b + n_t$ induced by either of the gates. This provides yet another proof that the observed QHE involves the entire graphite bulk. Colour scale in **a** and **b**: navy-to-yellow is from 46.3 to 47.2 fF. The Landau levels that originate from the 2D accumulation layers at the graphite surfaces are marked by arrows. **c,** Calculated density of states (navy-to-yellow, 0 to $100 \cdot 10^{12}$ cm$^{-2}$eV$^{-1}$) using the SWMC model complemented with the self-consistent Hartree analysis of the surface accumulation layer. **d,** Charge density $n_l$, and potential, $U_{layer}$ (insets) profiles along the $z$-direction, calculated using self-consistent Hartree analysis for a 14-layer-thick graphite, at 13 T for different $n_t$. The charges are introduced by electrostatic gating from one side. Red, black, blue and green curves correspond to contributions from the states in $m = 0$, $1$, $2^\pm$ and $3^\pm$ Landau bands, respectively (extended states for $m = 0$ and $1$, and evanescent states for $m = 2^\pm$ and $3^\pm$.

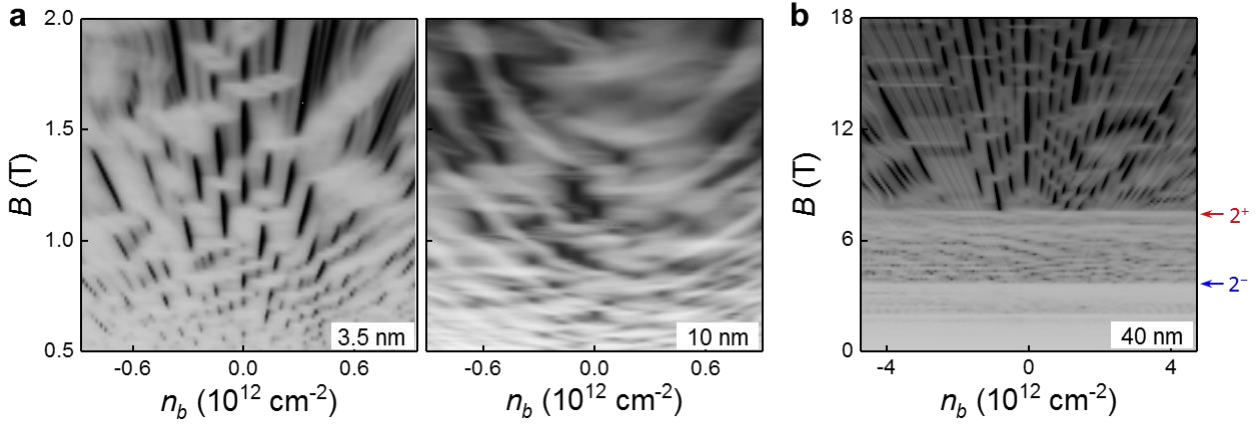

**Supplementary Fig. 4: Low-field QHE states and the ultra-quantum regime in graphite. a,** Conductivity maps $\sigma_{xx}(B, n)$ measured for 3.5 nm (left) and 10 nm (right) Corbino devices in low $B$. Black-to-white scale: 0 to 1 mS for the left panel and 0 to 0.7 mS for the right one. The QHE states (black regions) start developing already at 0.5 T. **b,** Conductivity map for a 40 nm thick Corbino device over the entire $B$ range. Logarithmic scale: black-to-white is 1 µS to 0.1 mS. The clearly-seen two horizontal boundaries (marked by the arrows) correspond to the fields of ∼ 3.7 and 7.6 T, in which bulk Landau bands *2⁻* and *2⁺*, respectively, become depopulated (see Fig. 2c of the main text). Above 7.6 T (red arrow) only the two lowest Landau bands (*0* and *1*) cross the Fermi level, which signifies the UQR for graphite. $T$ = 0.25 K for all the panels.

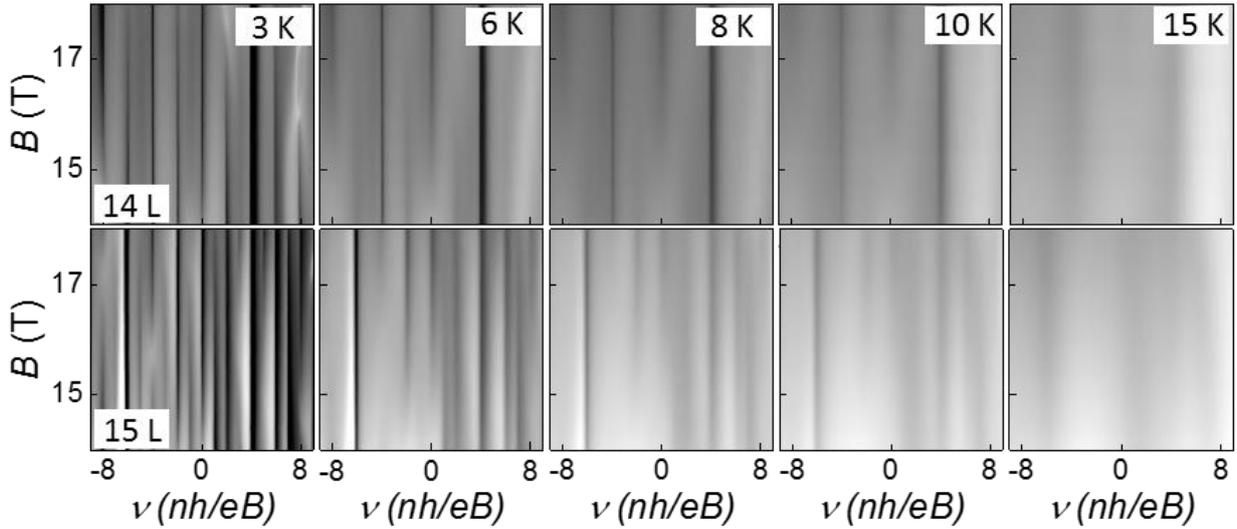

**Supplementary Fig. 5: Quantum Hall effect in even- and odd-layer graphite films.** Maps $\sigma_{xx}(B, \nu)$ were measured at different $T$ for devices with $N$ = 14 and 15 layers (top and bottom rows, respectively). The even-layer crystal exhibits the QHE with double degeneracy, reflected by the period of $\Delta\nu$ = 2, whereas $\Delta\nu$ = 1 for the odd-layer film, where all the degeneracies are lifted. Black-to-white scale is 0 to 1.0 µS and 0.2 to 1.3 µS for the top and bottom panels, respectively.

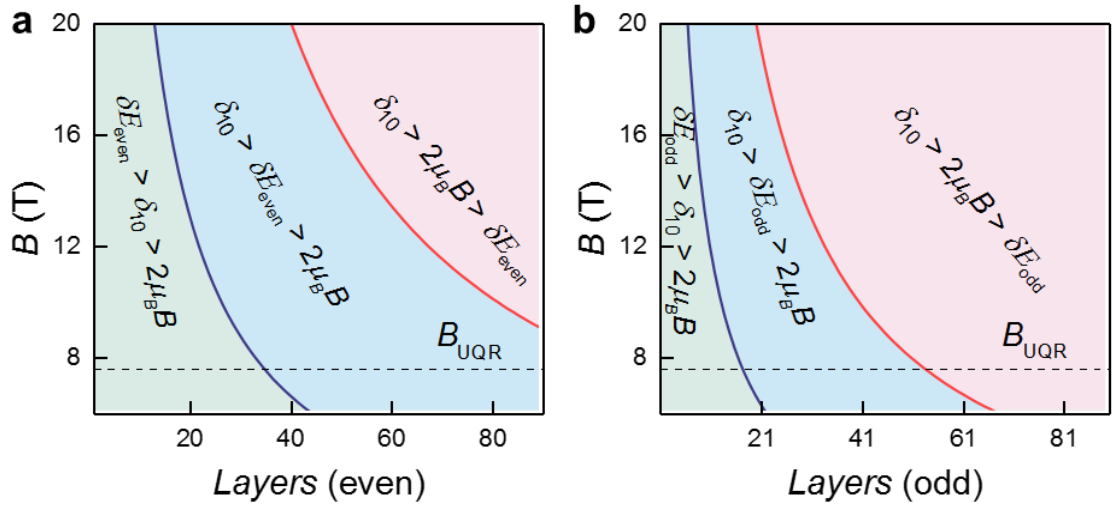

**Supplementary Fig. 6: Parameter space for graphite crystals with odd and even numbers of layers. a** and **b**, Different parameter regimes for even and odd *N*, respectively, as a function of the number of graphene layers and magnetic field. The standing wave spacing $\delta E_{even}$ (panel **a**) and $\delta E_{odd}$ (**b**) are compared with the Zeeman splitting $2\mu_B B$ and the splitting $\delta_{10}$ between the two lowest Landau bands. Dashed lines indicate the UQR boundary.

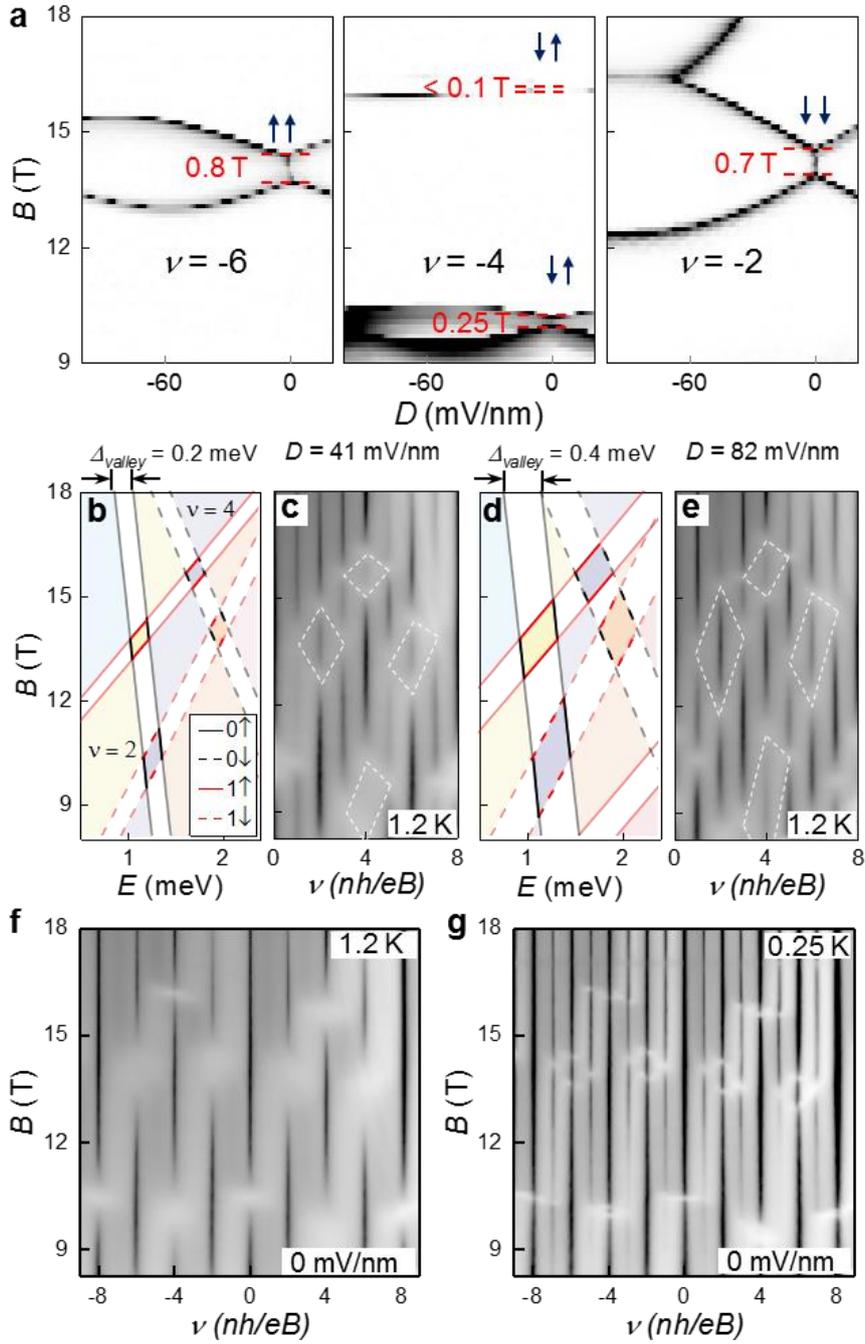

**Supplementary Fig. 7: Symmetry breaking in graphite crystals with even numbers of layers. a,** Level crossings for different spin-polarised states. Shown is $\sigma_{xx}$ as a function of $B$ and the displacement field $D = \Delta n \cdot e/2\varepsilon_0$ for the fixed filling factors $\nu = -6$ (left), $-4$ (middle) and $-2$ (right); $T = 0.25$ K. The white-to-black scales are 0 to 0.5 µS. The observed crossings exhibit quite different widths (highlighted by red dashed lines) depending whether they contain spins oriented in the same (↑↑) or opposite (↑↓) directions. **b-e,** Valley splitting in a 34-layer-thick graphite film. **b,** Standing wave spectra, calculated for valley splitting energy of 0.2 meV for *0* (black) and *1* (red) Landau bands. Solid (dashed) lines are for ↑ (↓) spin states. The coloured background specifies even filling factors: $\nu = 0$ (blue), 2 (yellow), 4 (violet), 6 (orange) and 8 (pink). **c,** Experimental map $\sigma_{xx}(B, \nu)$ for a 11.6 nm device (same as in Fig. 2 of the main text) at $D = 41$ mV/nm and $T = 1.2$ K. **d** and **e,** same as plots in **b** and **c** but using the valley splitting energy of 0.4 meV for the theory plot and $D = 82$ mV/nm in the measurements. Schematics of the valley-split bands in **b** and **d** highlight the diamond patterns that are also seen in the experimental data in **c** and **e** where they are indicated by the dashed white lines. **f, g,** Spontaneous valley polarisation. Conductivity $\sigma_{xx}(B, \nu)$ of the same 11.6 nm device at $D = 0$ at temperatures of 1.2 K (**f**) and 0.25 K (**g**). Logarithmic scales: black-to-white is 0.3 to 7.3 µS for **c, e**; 0.1 to 7.4 µS for **f**; 0.1 to 7.7 µS for **g**.

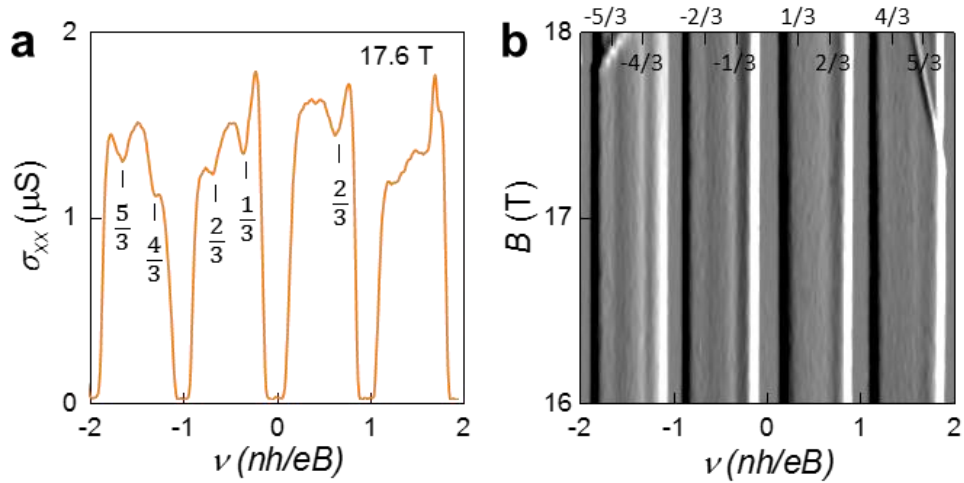

**Supplementary Fig. 8: Evidence for the fractional QHE in 3.5-nm-thick graphite. a,** $\sigma_{xx}$ as a function of $\nu$ measured in the Corbino geometry $T = 0.25$ K. **b,** The fractional features are clearly seen plotting differential conductivity d$\sigma_{xx}$/d$\nu$ as a function of $B$ and $\nu$.

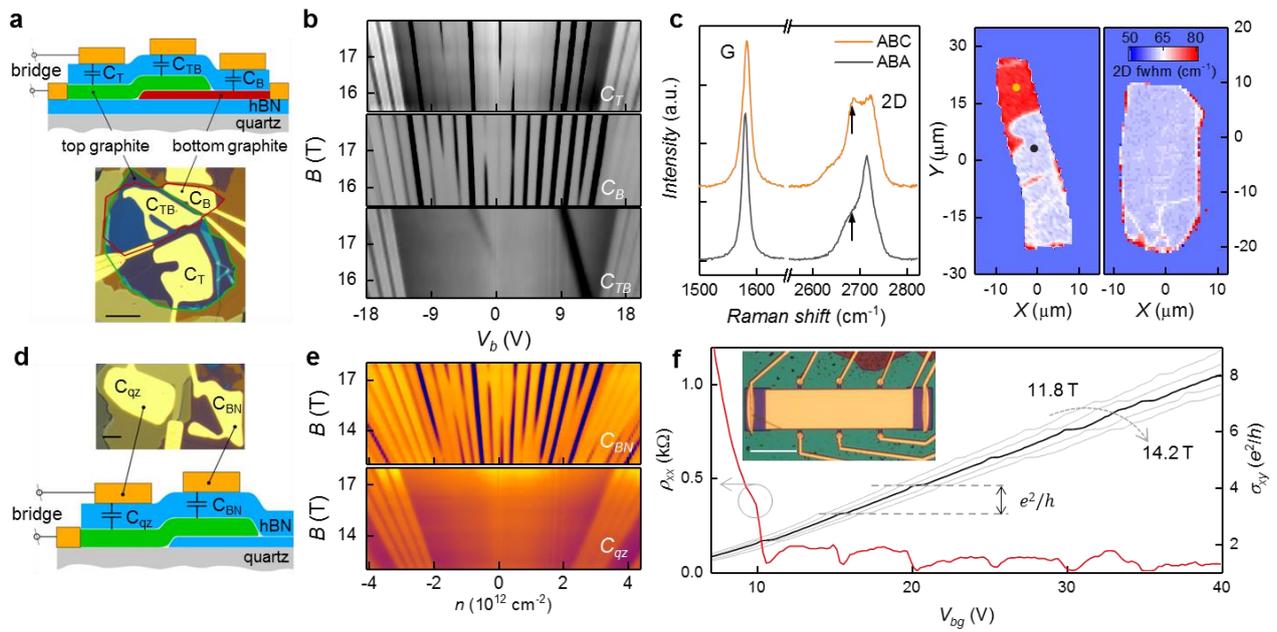

**Supplementary Fig. 9: How robust is the 2.5D QHE? a** and **b,** Effect of stacking faults. **a,** Schematic diagram (upper panel) and optical micrograph (lower) of a graphite device with an artificial stacking fault made by placing two graphite crystals on top of each other at a misalignment angle ≈ 2°. The top and bottom graphite are 9 nm and 5 nm thick, respectively. Scale bar in the lower panel, 10 μm. **b,** Capacitance as a function of $B$ and gate bias $V_b$ for the three regions: $C_T$ contains only the top graphite crystal, $C_B$ only the bottom graphite and $C_{TB}$ is the overlapping regions that represent the stacking fault. Black-to-white scales: 126 to 127.6 fF for $C_T$, 43.8 to 47.1 fF for $C_B$ and 80.5 to 82.0 fF for $C_{TB}$. $T$ = 0.25 K. **c,** Left panel: Typical Raman spectra of Bernal (ABA) graphite and graphite with ABC stacking. Middle: Raman map of a graphite flake with regions of both ABA and ABC staking. The colour coding follows the fwhm (full width at half maximum) of the 2D peak. The dots mark the positions where the Raman spectra shown in the left panel were taken. Right panel: Similar map for graphite crystals with pure ABA stacking. **d** and **e**, Effect of rough substrates. **d,** Schematic and optical micrograph of a capacitance device where a part of the same graphite crystal (5 nm thick) is encapsulated (marked as $C_{BN}$) whereas the other part ($C_{qz}$) is placed on a rough quartz substrate. Scale bar for the micrograph, 5 μm. **e,** Capacitance maps for these two parts at 0.25 K. Scale bars: navy-to-yellow, 66 to 74 fF for $C_{BN}$ and 98.2 to 101.4 fF for $C_{qz}$. The 2.5D QHE gaps are completely smeared for graphite on the rough substrate. **f,** Effect of the lateral size. The 2.5D QHE was observed in rather large Hall bar devices such as the one shown in the inset. Scale bar for the optical micrograph, 20 μm; graphite thickness, 12.3 nm. Hall conductivity $\sigma_{xy}$ (black curve) and longitudinal resistivity $\rho_{xx}$ (red) are shown as a function of back gate voltage $V_{bg}$ at $B$ = 13 T; $T$ = 0.25 K; top gate voltage $V_{tg}$ = 0. Grey curves: $\sigma_{xy}$ for $B$ around 13 T (from 11.8 to 14.2 T).